\documentclass[12pt]{article}
\usepackage{amsmath}
\usepackage{amssymb} 

\def\QED{\hskip0.1em\hfill\null\ \null\nobreak\hfill\kern3pt\vbox{\hrule\hbox
   {\vrule\kern1pt\vbox{\kern1.7pt\hbox{$\scriptscriptstyle{QED}$}
    \kern0.2pt}\kern1pt\vrule}\hrule}}

\def\END{\hskip0.1em\hfill\null\ \null\nobreak\hfill\kern3pt\vbox{\hrule\hbox
   {\vrule\kern1pt\vbox{\kern1.7pt\hbox{$\,\,\,\vspace{5pt}$}
    \kern0.2pt}\kern1pt\vrule}\hrule}}
\newtheorem{theorem}{Theorem}
\newtheorem{lemma}{Lemma}
\newtheorem{corollary}{Corollary}
\newtheorem{proposition}{Proposition}
\newtheorem{remark}{Remark}
\newtheorem{definition}{Definition}
\newtheorem{example}{Example}

\newcommand{\bCd}{\bEq\begin{CD}}
\newcommand{\eCd}{\end{CD}\eEq}
\newcommand{\bcd}{\beq\begin{CD}}
\newcommand{\ecd}{\end{CD}\eeq}
\newcommand{\ben}{\begin{enumerate}}
\newcommand{\een}{\end{enumerate}}
\newcommand{\bEq}{\begin{eqnarray}}
\newcommand{\eEq}{\end{eqnarray}}
\newcommand{\beq}{\begin{eqnarray*}}
\newcommand{\eeq}{\end{eqnarray*}}
\newcommand{\bDf}{\begin{definition}\em}
\newcommand{\eDf}{\end{definition}}
\newcommand{\bLm}{\begin{lemma}}
\newcommand{\eLm}{\end{lemma}}
\newcommand{\bPr}{\begin{proposition}}
\newcommand{\ePr}{\end{proposition}}
\newcommand{\bTh}{\begin{theorem}}
\newcommand{\eTh}{\end{theorem}}
\newcommand{\bCr}{\begin{corollary}}
\newcommand{\eCr}{\end{corollary}}
\newcommand{\bRm}{\begin{remark}\em}
\newcommand{\eRm}{\end{remark}}
\newcommand{\bEx}{\begin{example}\em}
\newcommand{\eEx}{\end{example}}
\newcommand{\C}{\mathbb{C}}

\newcommand{\ie}{{\em i.e$.$} }
\newcommand{\eg}{{\em e.g$.$} }

\newcommand{\A}{\forall}

\newcommand{\der}{\partial}

\newcommand{\bz}{\boldsymbol{z}}

\newcommand{\bE}{\boldsymbol{E}}

\newcommand{\bG}{\boldsymbol{G}}

\newcommand{\bK}{\boldsymbol{K}}

\newcommand{\bP}{\boldsymbol{P}}

\newcommand{\bV}{\boldsymbol{V}}

\newcommand{\bZ}{\boldsymbol{Z}}

\newcommand{\car}{\times}

\newcommand{\wed}{\wedge}

\newcommand{\gam}{\gamma}
\newcommand{\del}{\delta}

\newcommand{\tht}{\theta}

\newcommand{\lam}{\lambda}

\newcommand{\ome}{\omega}

\title{\large{{\bf Infinitesimal algebraic skeletons for a $(2+1)$-dimensional Toda type system}\thanks{Supported by 
the University of Torino through the local research project `Conservation laws in classical and quantum gravity'.
}}}
\author{{\normalsize M.
Palese and E. Winterroth
}\\ {\footnotesize Department of Mathematics,
University of Torino}
\\{\footnotesize via C. Alberto 10, 10123 Torino, Italy}\\  {\footnotesize e--mails: 
{\sc [marcella.palese, ekkehart.winterroth]@unito.it
}}}
\date{}
\begin{document}

\maketitle

\begin{abstract}
A tower for a $(2+1)$-dimensional Toda type system is constructed in terms of a series expansion of operators which can be interpreted as generalized Bessel coefficients; the result is formulated as an analog of the Baker--Campbell--Hausdorff formula. We tackle the problem of the construction of infinitesimal algebraic skeletons for such a tower and discuss some open problems arising along our approach. In particular, we realize the prolongation skeleton as a Ka\v{c}-Moody algebra.
\end{abstract}

\noindent {\bf Key words}: Toda type system, integrability, infinitesimal skeleton, tower, Cartan connection.

\noindent {\bf 2000 MSC}: 53C05, 53C80, 58A15, 58J70,  58J72, 70G65.
\section{Introduction}

Nonlinear models, and in particular Toda type systems, play a role in a variety of physical phenomena. As is well known, the problem of their integrability is far from being trivial. 
It is nowadays well recognized that the {\em algebraic properties} of nonlinear systems are relevant from the point of view of integrability. A huge scientific production within this topic has developed in both discrete and continuous, as well as, classical and quantistic models.
It is nevertheless important not to forget the origin of this interest: for a nonlinear system,
it lies in the concept of integrability as of having `enough' conservation laws to exaustively describe the dynamics (an idea which originates in the inverse of  
 the Noether Theorem II in the calculus of variations).
Historically, the algebraic-geometric  approach is based on the {\em requirement for the 
existence of conservation laws} 
which leads to the existence of symmetries (in terms of algebraic structures).

In this light, Wahlquist and Estabrook \cite{Wahlquist,Estabrook} proposed a  technique for systematically deriving what they called a `prolongation structure' in terms of a set of `pseudopotentials'
 related with the existence of an infinite set of associated conservation laws,
and they also conjectured that the structure was `open' \ie not a set of structure relations of a finite--dimensional Lie group. Since then, `open' Lie algebras have been extensively studied in order to distinguish them from freely generated infinite-dimensional Lie algebras.  

In their approach, conservation laws  are written in terms of `prolongation' forms and integrability is intended as an
integrability condition for a `prolonged' differential ideal.
Attempting a {\em description of symmetries in terms of Lie algebras} implies the appearance of an homogeneous space and thus the interpretation of prolongation forms as {\em Cartan--Ehresmann connections}.
It should be stressed that the unknowns are both conservation laws and symmetries,
and it is clear that the main point in this is how to realize the form of the conservation laws 
and thus the explicit expression of the prolongation forms.
Different formulations of the prolongation ideal bring to both different algebraic structures (symmetries) and corresponding conservation laws: of course,
 the structure with which prolongation forms are postulated can produce Lie algebras or more general algebraic structures.
We use the algebraic properties of Toda type systems as a `laboratory' to explicate 
an algebraic-geometric interpretation of the above mentioned `prolongation' procedure 
in terms of towers with infinitesimal algebraic skeletons \cite{Morimoto}.

Consider the $(2+1)$-dimensional system, a continuous (or long-wave) approximation of a spatially two-dimensional Toda lattice \cite{Toda}:
\bEq\label{1}
u_{xx}+u_{yy}+(e^{u})_{zz}=0\,,
\eEq
where $u=u(x,y,z)$ is a real field, $x,y,z$ are real local coordinates (if we want, $z$ playing the r\^ole of a `time') and the subscripts mean partial derivatives.
  
It can be seen as the limit for $\gam\to \infty$ of the more general model 
$u_{xx}+u_{yy}+\left[\left(1+u/ \gam \right)^{\gam -1}\right]_{zz}=0$, covering (for $\gam$ different from $0,1$) various continuous approximations of  lattice models, among them the Fermi-Pasta-Ulam ($\gam =3$).
 It appears in differential geometry:
 Kaehler  metrics \cite{Lebrun} ;
    in mathematical and theoretical physics (see, \eg  Newman and Penrose as well as \cite{ 
    Plebanski});
in the theory of Hamiltonian systems, in general relativity: heavenly spaces (real, self-dual, euclidean Einstein spaces with one rotational Killing symmetry, \cite{Plebanski,Boyer});
   in the large $n$ limit of the $\textstyle{sl}(n)$ Toda lattice \cite{Park} (from the constrained Wess-Zumino-Novikov-Witten model):
           extended conformal 
symmetries ($2D$  $CFT$) and reductions of  four dimensional theory of gravitational instantons;
in strings theory and statistical mechanics.
It can be seen as the particular case with $d=1$ of so-called  $2d$-dimensional 
Toda-type systems \cite{Saveliev} from a `continuum Lie algebra' by means of a zero curvature representation $u_{w\bar{w}}= K(e^{u})$, (in our particular case $w =x+iy$ and $K$ is  the differential operator  given by $K= \frac{\der^{2}}{\der z^{2}}$). 
In particular, it has been studied in the context of symmetry reductions \cite{Alfinito1, Grassi} and a $(1+1)$-dimensional version in the context of prolongation structures which only partially lead to results  \cite{Alfinito2}.

\section{Towers with skeletons for Toda type systems}

The notion of an (infinitesimal) algebraic skeleton is an abstraction
of some algebraic aspects of homogeneous spaces.
Let then $\bV$ denote a finite--dimensional vector space. 

An {\em algebraic skeleton} on $\bV$ is a triple $(\bE,\bG,\rho)$, with 
$\bG$ a
(possibly infinite-dimensional) Lie group, 
$\bE=\bV\oplus\mathfrak{g}$, 
$\mathfrak{g}$ the Lie
algebra of $\bG$, and 
$\rho$ a representation of $\bG$ on $\bE$ (infinitesimally of $\mathfrak{g}$ on $\bE$) such that
$\rho(g)x=Ad(g)x$, for $g\in\bG$, $x\in\mathfrak{g}$.

Let  $\bZ$ be a manifold of type
$\bV$ (\ie  $\A \bz\in \bZ$,  $T_{\bz} \bZ \simeq \bV$).
We say that a  principal fibre bundle
$\bP(\bZ,\bG)$ provided with an 
{\em  absolute parallelism} $\omega$ on $\bP$
is a {\em tower} on $\bZ$ with skeleton $(\bE, \bG,\rho)$
 if $\omega$ takes values in $\bE$
and satisfies: 
 $R^{*}_{g}\omega = \rho(g)^{-1}\ome$, for
$g\in \bG$; $\ome(\tilde{A}) = A$, for $A\in\mathfrak{g}$;
here $R_{g}$ denotes the right translation and $\tilde{A}$ the fundamental
vector field induced on $\bP$ from $A$.
In general, the absolute parallelism  {\em does not}  define a Lie algebra homomorphism.

Let $\mathfrak{g}$ be a Lie algebra and $\mathfrak{k}$ be a Lie subalgebra of
$\mathfrak{g}$. Let $\bK$ be a Lie group with Lie algebra $\mathfrak{k}$ and let $\bP(\bZ, \bK)$ be a principal fibre bundle with structure group $\bK$ over a manifold $\bZ$, as above. 
A {\em Cartan connection} in $\bP$ of type
$(\mathfrak{g}, \bK)$ is 
a $1$--form $\omega$ on $\bP$ with values in
$\mathfrak{g}$ satisfying the following conditions: 
- $\omega|_{T_{u} \bP}: T_{u} \bP\to \mathfrak{g}$ is an isomorphism $\forall u\in
\bP$; 
- $R^{*}_{g}\omega=Ad(g)^{-1}\omega$ for $g\in \bK$;
- $\omega(\tilde{A})=A$ for $A \in \mathfrak{k}$.
A  Cartan connection
$(\bP, \bZ, \bK, \omega)$ of type $(\mathfrak{g}, \bK)$ is a tower on $\bZ$.

Remark that since, a priori, the prolongation algebra does not close into a Lie algebra the starting point for the prolongation procedure is only a tower with an absolute parallelism, and not a Cartan connection.
Thus, in principle, Estabrook-Wahlquist {\em prolongation forms are  absolute parallelism forms}.
The corresponding open Lie algebra  structure can be provided with the 
structure of an infinitesimal algebraic skeleton on a suitable space.
First we have to prove that a finite dimensional space $\bV$ and a Lie algebra $\mathfrak{g}$ exist satisfying the definition of a skeleton, \ie in particular that a suitable representation $\rho$ can be defined.
The representation is obtained by means of an integrability  
condition for the {\em absolute parallelism} of a tower on a manifold $\bZ$ (of type $\bV$), with skeleton  $(\bE, \bV, \mathfrak{g})$. 

Note that {\em if} $\bE$ has in addition the structure of a Lie algebra this is exactly a Cartan connection of type $(\bE, \mathfrak{g})$; in fact, the spectral linear problem is nothing but 
the construction of a Cartan connection from this absolute parallelism.

As an example, let us now introduce on a manifold with local coordinates
$(x,y,z,u,p,q,r)$ the closed differential ideal defined by the set of $3$--forms:
$
\theta_{1} = du \wedge dx \wedge dy - rdx \wedge dy \wedge 
dz$, 
$\theta_{2} = du \wedge dy \wedge dz - pdx \wedge dy \wedge 
dz$,
$\theta_{3} = du \wedge dx \wedge dz + qdx \wedge dy \wedge 
dz$,
$\theta_{4} = dp \wedge dy \wedge dz - dq \wedge dx \wedge 
dz$ $ +$ $ e^{u}dr\wedge dx \wedge dy + e^{u}r^{2}dx \wedge dy
\wedge dz$. 
It is easy to verify that on every integral submanifold defined by $u = u(x,y,z)$, 
$p=u_{x}$, $q=u_{y}$, $r=u_{z}$, with $dx\wedge dy\wedge dz\neq 0$, the above ideal
is equivalent to the Toda system under study. 

In terms of absolute parallelism  forms, $2$--forms generating associated conservation laws can be defined as follows:
\beq \Omega^{k} = H^{k}(u, u_{x}, u_{y}, u_{z}; \xi^{m})dx \wedge dy+
F^{k}(u, u_{x},u_{y}, u_{z}; \xi^{m})dx \wedge dz + \\
+ G^{k}(u, u_{x}, u_{y}, u_{z}; \xi^{m})dy \wedge dz+ A^{k}_{m}d \xi^{m} \wedge dx
+ B^{k}_{m}d \xi^{m} \wedge dz + d \xi^{k}
\wedge dy\,,
\eeq 
where $\xi=\{\xi^{m}\}$, $k,m=1,2,\ldots,
{\rm N}$ (N arbitrary), and $H^{k}$, $F^{k}$ and $G^{k}$ are, respectively, the
pseudopotential and functions to be determined, while $A^{k}_{m}$
and $B^{k}_{m}$ denote the elements of two $N \times N$ 
constant regular matrices.
In fact, we remark that $\Omega^{k}= \tht^k_m\wed\ome^m$, where 
$\tht^k_m = - \bar{A}^{k}_{m} dx-\bar{B}^{k}_{m}dy- \bar{C}^{k}_{m}dz$,  and the absolute parallelism forms are given by\footnote{$F^k = \bar{C}^{k}_{m}\bar{F}^m- \bar{A}^{k}_{m}  \bar{H}^m$, 
$G^k = \bar{C}^{k}_{m}\bar{G}^m- \bar{B}^{k}_{m}  \bar{H}^m$,
$H^k = \bar{B}^{k}_{m}\bar{F}^m- \bar{A}^{k}_{m}  \bar{G}^m$, $\xi^k = \bar{C}^{k}_{m}\bar{\xi}^m$} 
\beq
\ome^m= d\bar{\xi}^m+ \bar{F}^m dx  +\bar{G}^m dy  +  \bar{H}^m dz\,.\eeq

The integrability  condition for the ideal generated by forms 
$\theta_{j}$ and $\Omega^{k}$ finally yields
$H^{k} $ $=$ $ e^{u}u_{z}L^{k}(\xi^{m}) + P^{k}(u,\xi^{m})$,
$F^{k} $ $=$ $ -\,u_{y}L^{k}(\xi^{m}) + N^{k}(\xi^{m})$,
$G^{k}$ $=$ $ u_{x}L^{k}(\xi^{m}) + M^{k}(u,\xi^{m})$,
where $L^{k}$, $P^{k}$, $N^{k}$, $M^{k}$ are functions of integration. 
As a consequence, the desired  representation for the skeleton is provided by the following equations (we omit the indices for simplicity).
\bEq
P_{u} = e^{u}[L,M]\,,\label{P} \, \quad
 M_{u} = - [L,P]\,,\label{M} \, \quad 
 \left[M,P\right] = 0\,.\label{comm}
\eEq

We will consider $L$, $P$, $M$ as regular operators so that Lie brackets can be interpreted as commutators.
We can now look for an exact solution in order to give the representation explicitly. 
For any operator $D = D^{j}\frac{\partial}{\partial \xi^{j}}$, by introducing  ${\cal{L}}[D] = [L, D]$, 
we define the $n$--th power of the operator ${\cal{L}}$ by setting
${\cal{L}}^{n} [D] = [L,[L,\dots ,[L, D]\dots]$, where $L$ appears $n$--times, and ${\cal L}^{0}[D]$ $=$ $D$.

Put $t = 2e^{\frac{u}{2}}$.
A solution of the prolongation equations regular at $t=0$ ({\em i.e.} at $u\to -\infty$) is then given by
\bEq 
P = \frac{t}{2}{\bf J_1} (t{\cal{L}} [P_0])\,,\qquad
M = {\bf J_0} (t{\cal{L}} [M_0])\,,\label{first solution}
\eEq 
where 
${\bf J_0} (\cdot)$ and ${\bf J_1} (\cdot)$are formal operator expansions given by 
${\bf J_0} (t{\cal{L}}[M_0]) = 
\sum_{m=0}^{\infty}\frac{(-1)^m}{(m!)^2}
\left(\frac{t}{2}\right)^{2m}{\cal{L}}^{2m}
[M_0]$,
${\bf J_1} (t{\cal{L}} [P_0]) =
\sum_{m=0}^{\infty}\frac{(-1)^m}{m!(m+1)!}
\left(\frac{t}{2}\right)^{1+2m}{\cal{L}}^{1+2m}
[P_0]$ and $M_0 \equiv M_0(\xi) = M(t;\xi)\mid _{t=0}$ and
$P_0 \equiv P_0(\xi)$ is such that $[L,P_0] = [L,M_0]$ \cite{Palese}.

By defining {\em operator Bessel coefficients} ${\bf J_m}(tX)$, 
as the coefficients of the formal expansion
$
e^{\frac{t}{2}X(z-1/z)} = \sum_{m=-\infty}^{\infty}z^{m}{\bf J_{m}}(tX)$ (for Bessel functions a standard reference is \cite{Watson}), 
we can prove recurrence and derivation formulae
by means of which we provide an equivalent
solution to our prolongation equations in terms of $L$:
\beq
P = \frac{t}{2}\sum_{k=-\infty}^{\infty}{\bf J_{k+1}}(tL)P_{0}{\bf 
J_{k}}(tL)\,, \,
M = \sum_{k=-\infty}^{\infty}{\bf J_{k}}(tL)M_{0}{\bf 
J_{k}}(tL)\,,
\eeq
based on the formulae 
${\bf J_1} (t{\cal{L}} [P_0])=\sum_{k=-\infty}^{\infty}{\bf J_{k+1}}(tL)P_{0}{\bf 
J_{k}}(tL)$,
${\bf J_0} (t{\cal{L}} [M_0])= \sum_{k=-\infty}^{\infty}{\bf J_{k}}(tL)M_{0}{\bf 
J_{k}}(tL)$,
which are in fact analogous  to the Baker--Campbell--Hausdorff expansion \cite{Palese}. These expansions together with $\left[M,P\right] = 0$ provide the desired representation and at the same time define a tower with absolute parallelism.

The main problem with this tower (which is somehow the most general one) is that it is a non trivial task to characterize explicitly its algebraic skeleton by means of the representation provided by  the relations $\left[M,P\right] = 0$.
On the other hand, it is well known that the Toda equation can be solved  by the inverse scattering transform 
\cite{ Kodama}. However, the associated linear spectral problem was never derived from an infinitesimal algebraic skeleton and in particular as the construction of a Cartan connection from a tower with algebraic skeleton; thus it would be important to derive both the Toda system and related spectral problem(s) (\ie conservation laws and symmetries) starting from a tower with an algebraic skeleton.
In this perspective, particular solutions of the corresponding Estabrook-Wahlquist prolongation problem can assume a relevant role: they correspond to particular choices for the absolute parallelism and can provide us explicit representations of the prolongation skeleton. 

\subsection{Skeletons}

If we look for operators $P(u,\xi)$ and $M(u,\xi)$ depending on $u$ only through the exponential function, \ie
$P(u,\xi)= e^{u}\bar{P}(\xi)$, $M(u,\xi)=M(e^u,\xi)$, the prolongation equations can now be written as:
$P_{u}= e^u [L,M] = \frac{\der P}{\der e^u}e^u $,
$M_{u} = - [L,P] = \frac{\der M}{\der e^u}e^u $; on the other hand, we have 
$\frac{\der P}{\der e^u}$ $=$ $\bar{P}(\xi)$ $=$ $[L(\xi),M(e^u;\xi)]$, 
$\frac{\der M}{\der e^u} $ $=$ $ -[L(\xi),\bar{P}(\xi)]$.

\noindent From the second equation, we get
$M( e^u;\xi) $ $=$ $ - e^u[L(\xi),\bar{P}(\xi)]+ \bar{M}(\xi)$ and thus
$\bar{P}(\xi)$ $=$ $ - e^u[L(\xi),[L(\xi),\bar{P}(\xi)]]+[L(\xi),\bar{M}(\xi)]$.

\noindent We see then that we are able to obtain commutation relations:
$\bar{P}(\xi)
 =[L(\xi),\bar{M}(\xi)]\,, \quad [L(\xi),[L(\xi),\bar{P}(\xi)]]=0$.
There are additional relations determined by the third prolongation equation 
$[- e^u[L(\xi),\bar{P}(\xi)]+ \bar{M}(\xi), e^u \bar{P}(\xi)]=0$, so that we have 
$[[L,\bar{P}],\bar{P}]$ $=$ $0$, $[\bar{M},  \bar{P}]$ $=$ $0$.
For the sake of convenience we put $L=X_1,\bar{M}=X_2, \bar{P}=X_3, [X_1 ,X_3 ]= X_4 $ and we then have the following prolongation {\em closed Lie algebra}:
\beq
[X_1 ,X_2 ]=X_3 \,,   [X_1 ,X_3 ]= X_4 \,  [X_1 , X_4 ]= [X_2 ,X_3 ]= [X_2 ,X_4 ]= [X_3 ,X_4 ]= 0 \,.
\eeq

Suppose now that
$P(u,\xi)= \ln u \bar{P}(\xi)$, $M(u,\xi)=M(e^u,\xi)$.
We derive then 
$P_{u}$ $=$ $ e^u [L,M] $ $=$ $ \frac{d  (\ln u \bar{P}(\xi))}{d e^u}e^u$ $=$ $ \frac{1}{u}\bar{P}(\xi)$,
$M_{u} $ $=$ $ \frac{\der M}{\der e^u}e^u $ $=$ $ - [L,P] $ $=$ $ - [L,\ln u \bar{P}(\xi)]$; so that 
$\frac{\der M}{\der e^u}$ $=$ $ - \frac{\ln u }{e^u}[L,\bar{P}(\xi)]$, from which we get
$M(e^u,\xi)=- (\ln u -1)u[L(\xi),\bar{P}(\xi)]+\bar{M} (\xi)$,
and 
$P(u,\xi)= ue^u \ln u[L , M] $.
From $[P,M]=0$
we get, for $u \neq 0,1$ (which are trivial solutions of the Toda system),
\beq
[[L,M],M]=0\,;
\eeq
on the other hand substituting the above expression for $M$
we get 
\beq
[ [L,\bar{M}],\bar{M} ] = 0 \,, 
\,
[ [L,[L,\bar{P}] ], \bar{M} ]+ [ [L,\bar{M}], [L,\bar{P}] ] =0\,,
\,
[ [L, [L,\bar{P}] ], [L,\bar{P}] ] = 0\,.
\eeq
by putting again for the sake of convenience $L=X_1,\bar{M}=X_2, \bar{P}=X_3$,
then we get the following infinitesimal algebraic  skeleton with the structure of an {\em open Lie algebra}:
\beq
[X_1, X_2]=X_4\,, [X_1, X_3]=X_5\,, [X_4, X_5]  
=[X_2,X_7] \,, [X_3, X_4] =  [X_2, X_5] \,,
\eeq
\beq
 [X_1, X_4]=X_6\,,  [X_1, X_5]=X_7\,, [X_2, X_3] = X_8  \,, 
\eeq
\beq
[X_1, X_8]= [X_2, X_4] =  [X_2, X_6] =[X_3, X_7] = 0\,,\\
\ldots
\eeq
We observe that by the homomorphism $X_4=X_5=0$ and $X_8=\nu X_3$ we get a {\em closed Lie algebra}:
\beq
[X_1, X_2]=0\,, [X_1, X_3]=0\,, [X_2, X_3] = \nu X_3\,.
\eeq
which, by means of a suitable 
realization, can also provide us  with a different Cartan connection (thus a different spectral problem and different conservation laws);
on the other hand, we can find a {\em closed Lie algebra}  by means of the following homomorphism $X_4=X_2$ and $X_5 =X_3$, and then we have 
\beq
[X_1, X_2]=X_2\,, [X_1, X_3]=X_3\,, [X_4, X_5] = [X_2, X_3] = X_8 =0
 \,,
\eeq
\beq 
[X_3, X_4] =  [X_3, X_2]=- [X_2, X_3] =  [X_2, X_5] = [X_2, X_3]  \,,
\eeq
and we also deduce that $X_6=X_4=X_2$, $X_7=X_3$ and that  $[X_1, X_8]$ $=$ $[X_2, X_4]$ $=$ $ [X_2, X_6]$ 
$=$ $[X_3, X_7]$ $=$  $0$ are all identically satisfied.

It is easy to see that the two different cases above are both given by the homomorphism given by requiring $X_4=\lam X_2$ and $X_5 =\mu X_3$. 
It is easy to realize that $\mu=-\lam$ must old and there are the two cases $\lam =0$ with $X_8=\nu X_3$ giving the first case, and $X_8=0$ with $\lam=1$ giving the second case, respectively.

For any $\lam \neq 0$ we have a closed Lie algebra depending on the parameter  $\lam$:
\beq
[X_1, X_2] = \lam X_2\,, [X_1, X_3] = - \lam X_3\,, [X_2, X_3] = 0 \,.
\eeq

Furthermore, by putting in the prolongation skeleton $X_4= X_2$ and $X_5 =- X_3$ it is possible to realize the prolongation skeleton as a Ka\v{c}-Moody Lie algebra of the type
\beq
 [h_i, h_j]=0\,, [h_i, X_{+ j}]= \kappa_{ij} X_{+ j} \,, \eeq
 \beq
 [h_i, X_{- j}]= -\kappa_{ij} X_{- j}\,, [X_{+ i}, X_{- j}]=\del_{ij} h_i\,,
 \eeq
where we put $[X_2, X_3] = X_8$, $[X_8, X_2] = X_9$, $[X_8, X_3] = X_{10}$, $[X_8, X_9] = X_{11}$ $[X_8, X_{10}] = X_{12}$ and 
$\{X_1,  X_{13}, \ldots \}=h_i$, $\{X_8, \ldots \}= h_j$, $\{X_2, X_9, X_{11}, \ldots \}= X_{+i}$, $\{X_3, X_{10}, X_{12}, \ldots \}= X_{-j}$. 

We also put $[X_8, X_{11}] =  X_{11}$, $[X_8, X_{12}] =- X_{12}$, and so on.
We then also  have $[X_8, X_{13}] =0$ and 
it is easy to realize that $[X_1, X_9] = X_9$, $[X_1, X_{10}] =- X_{10}$, $[X_1, X_{11}] = X_{11}$, $[X_1, X_{12}] = - X_{12}$, and so on; thus characterizing the Cartan matrix $\kappa_{ij}$.

It would be of interest to study the relation of skeletons with generalization of continuum Lie algebras to the case when the local algebra does not  generate $\mathfrak{g}(E; K, S)$ as a whole, where  $\mathfrak{g}(E; K, S)$ are
Saveliev's continuum Lie algebras and they are defined as follows.
Let $E$ be  a vector space parametrizing Lie algebras $\mathfrak{g}_i$, $i=0, +1, -1$, $\hat{\mathfrak{g}}\equiv \mathfrak{g}_{-1}\oplus \mathfrak{g}_0 \oplus \mathfrak{g}_{+ 1}$, such that
$
[ X_0 (\phi), X_0 (\psi) ]=0 \,, [ X_{+1}(\phi), X_{-1}(\psi) ]$ $=$ $X_0 (S(\phi, \psi))$,
$
[ X_0 (\phi), X_{+1} (\psi) ]$ $=$ $X_{+1}(K(\phi, \psi))$,
$
[ X_0 (\phi), X_{-1}(\psi) ]$ $=$ $- X_{-1}(K(\phi, \psi))
$, with $K$, $S$ bilinear maps $E\car E \to E$ satisfying conditions equivalent to the Jacobi identity.
Take $\mathfrak{g}' (E; K, S)$ as the Lie algebra freely generated  by a local part $\hat{\mathfrak{g}}$ and then the quotient 
$\mathfrak{g}(E; K, S)= \mathfrak{g}' (E; K, S)/ J$, $J$ the largest homogeneous ideal having a trivial intersection with $\mathfrak{g}_0$.
In fact, such an algebra becomes the Ka\v{c}-Moody  algebra above  when $E=\C^n$, $K=$ Cartan matrix $k$, $S=I$.
The relation with the Virasoro algebra without a central charge could be also considered in this light. This topic will be the object of further investigations.


\begin{thebibliography}{99}

\bibitem {Alfinito1} 
Alfinito, E., Soliani, G., and Solombrino, L.: 
{\em The symmetry structure of the heavenly equation}
 Lett. Math. Phys. {\bf 41}, 379-389 (1997).

\bibitem {Alfinito2} 
Alfinito, E.; Causo, M. S.; Profilo, G.; Soliani, G.: {\em A class of nonlinear wave equations containing the continuous Toda case}, J. Phys. {\bf A 31} (9) (1998)2173--2189.

\bibitem {Baker} 
Baker, H.F.:
{\em Alternant and continuous group}, Proc. London Math. Soc. (2), {\bf 3}, 24--47 (1904); 
Campbell, J.E.: {\em On a law of combination of operators}, Proc. London Math. Soc. {\bf 29}, 14--32 (1898); 
Hausdorff, F.: 
{\em The symbolic exponential formula in group theory}, Ber. Verh.S\"achs. Gess. Wiss. Leipzig. Math.--Phys. Kl. {\bf 58}, 
19--48 (1906).

\bibitem {Boyer} 
Boyer, C. and Finley, J.D.: 
{\em Killing vectors in self-dual, Euclidean Einstein spaces}, J. Math. Phys. {\bf 23}, 1126-1128 (1982).

\bibitem {Estabrook}
Estabrook, F.B. and Wahlquist, H.D.: 
{\em Prolongation structures of nonlinear evolution equations. II}, J. Math. Phys. {\bf 17}, 1293--1297 (1976).

\bibitem {Grassi} 
Grassi, V.; Leo, R. A.; Soliani, G.; Solombrino, L.: {\em Continuous approximation of binomial lattices}, Internat. J. Modern Phys. {\bf A 14} (15) (1999) 2357--2384. 

\bibitem {Kodama} 
Kodama, Y.: {\em Solutions of the dispersionless Toda equation},  Phys. Lett. {\bf A 147} (8-9) (1990) 477--482.

\bibitem {Lebrun} 
Lebrun, C.: {\em Explicit self-dual metrics on $CP_{2}\#...\#CP_{2},$}, J. Diff. Geom. {\bf 34} (1) (1991) 223-253 .

\bibitem {Morimoto} Morimoto, T.:
{\em Geometric structures on filtered manifolds}, Hokkaido Math. J. {\bf 22}(3) (1993) 263--347.

\bibitem {Palese}
 Palese, M.; Leo, R. A.; Soliani, G.: {\em The prolongation problem for the heavenly equation}; in {\em Recent developments in general relativity (Bari, 1998)} Springer Italia, Milan (2000) 337--344.
 
\bibitem {Park} 
Park, Q-Han: {\em Extended conformal symmetries in real heavens}, Phys. Lett. {\bf 236B}, 429-432 (1990).

\bibitem {Plebanski} 
Plebanski, J.F.: {\em Some solutions of complex Einstein equations}, J. Math. Phys. {\bf 16}, 2395--2402 (1975).

\bibitem {Saveliev}

Saveliev, M. V.:
{\em Integro-differential nonlinear equations and continual Lie algebras}, Comm. Math. Phys. {\bf 121} (2)(1989) 283--290; 
Saveliev, M. V.: {\em On the integrability problem of a continuous Toda system},  Theoret. and Math. Phys.(1992) {\bf 92} (3) 1024--1031 (1993);
Razumov, A. V.; Saveliev, M. V.: {\em Multidimensional systems of Toda type}, Theoret. and Math. Phys. (1997) {\bf 112} (2) 999--1022 (1998).

\bibitem {Toda} 
Toda, M.:  {\em Theory of nonlinear lattices}, Springer Series in Solid-State Sciences {\bf 20} Springer-Verlag, Berlin-New York (1981).

\bibitem {Wahlquist}
Whalquist, H.D. and Estabrook, F.B.: 
{\em Prolongation structures of nonlinear evolution equations}, J. Math. Phys. {\bf 16}, 1--7 (1975).

\bibitem {Watson} 
Watson, G.N.: {\em A Treatise on the Theory of Bessel 
Functions}, Cambridge University Press, Cambridge (1952).

\end{thebibliography}
\end{document}